\documentclass[prd,twocolumn,showpacs]{revtex4}
\usepackage{epsfig}
\usepackage{color}
\usepackage{amssymb}
\usepackage{amsfonts}
\usepackage{graphicx}
\usepackage{keyval,graphicx}
\usepackage{textcomp,wasysym}
\usepackage{subfigure}
\usepackage{slashed}
\usepackage{mathrsfs}

\newcommand{\be}{  \begin{eqnarray}}
\newcommand{\ee}{\end{eqnarray}}

\begin{document}
\title{Novel insights into the $\gamma\gamma^*\to \pi^0$ transition form factor}
\author{Ze-kun Guo$^1$\footnote {E-mail: guozk@ihep.ac.cn}
and Qiang Zhao$^{1,2}$\footnote {E-mail: zhaoq@ihep.ac.cn}}

\affiliation{1) Institute of High Energy Physics, Chinese Academy of
Sciences, Beijing 100049, P.R. China \\
2) Theoretical Physics Center for Science Facilities, CAS, Beijing
100049, China}

\begin{abstract}

BaBar's observation of significant deviations of the pion transition
form factor (TFF) from the asymptotic expectation with $Q^2>9$
GeV$^2$ has brought a serious crisis to a fundamental picture
established for such a simplest $q\bar{q}$ system by perturbative
QCD, i.e. the dominance of collinear factorization at high momentum
transfers for the pion TFF. We show that non-factorizable
contributions due to open flavors in $\gamma\gamma^*\to\pi^0$ could
be an important source that contaminates the pQCD asymptotic limit
and causes such deviations with $Q^2>9$ GeV$^2$. Within an effective
Lagrangian approach, the non-factorizable amplitudes can be related
to intermediate hadron loops, i.e. $K^{(*)}$ and $D^{(*)}$ etc, and
their corrections to the $\pi^0$ and $\eta$ TFFs can be estimated.

\end{abstract}

\date{\today}




\maketitle

\section{Introduction}

Since the foundation in late 1970s \cite{ERBL}, perturbative QCD
(pQCD) has been a powerful tool to explore the strong interaction
phenomena at large momentum transfers. This is such a kinematic
region where the exclusive transition matrix element can be
expressed as the convolution of the perturbative calculable
coefficient and non-perturbative light-cone distribution amplitude
(DA) that describes the longitudinal momentum fraction of quarks,
namely the so-called collinear factorization. In particular, with
the availability of clean electromagnetic probes, the photon-pion
transition form factor (TFF) will be the most ideal subject to test
the validity of pQCD and partonic structure of pion.

Experimentally, the pseudoscalar mesons' TFFs have been measured by
CELLO and CLEO collaborations with the virtuality of one photon
$Q^2$ up to 9 GeV$^2$ and the other nearly
on-shell~\cite{Behrend:1990sr,Gronberg:1997fj}. The data fit very
well the pQCD leading order interpolation
formula~\cite{Brodsky:1981rp}, which stitches together the chiral
anomaly and pQCD asymptotic limit,
\begin{eqnarray}\label{eq:inter}
F_{BL}^{\gamma
P}(Q^2)=\frac{1}{4\pi^2f_P}\frac{1}{1+(Q^2/8\pi^2f_P^2)},
\end{eqnarray}
where $f_\pi=92.3$ MeV, $f_\eta=97.5$ MeV ~\cite{Gronberg:1997fj}
are the corresponding decay constants of the pseudoscalar mesons. In
fact, it has been shown that nearly all the analyses up to
next-to-leading order (NLO) radiative and twist-four corrections
favor the endpoint-suppressed or asymptotic-like pion DA, which
means that the first two nontrivial Gegenbauer moments of the DA are
small and have opposite signs in the region of $Q^2=9$
GeV$^2$~\cite{Guo:2008zzg,Mikhailov:2009kf}.

However, the celebrated confirmation of pQCD dominance at large
momentum transfers was shortly overwhelmed by the surprising BaBar
measurement~\cite{Aubert:2009mc,:2011hk}. It shows that the rescaled
pion TFF exhibits a continuous growth at $9<Q^2 <40 \
\mathrm{GeV}^2$, even beyond the asymptotic limit of pQCD. In
contrast, the $\eta$ and $\eta^\prime$ TFFs seem to be consistent
with the pQCD expectations. Thus, a coherent understanding of the
$\pi^0$ and $\eta \ (\eta^\prime)$ TFFs would be a necessity for
resolving this puzzle.

The BaBar observation immediately arouses tremendous interests in
the pion TFF. In the pQCD scheme, the scaling violation of the BaBar
data can be better explained by either a wider DA with deep midpoint
or even a flat
one~\cite{Radyushkin:2009zg,Polyakov:2009je,Chernyak:2009dj,Li:2009pr,Wu:2010zc}.
However, they both conflict with some other constraints such as the
data at low $Q^2$, midpoint value of the pion DA from light-cone sum
rules~\cite{Agaev:2010aq}, or operator product expansion (OPE)
analysis for the leading handbag
contribution~\cite{Radyushkin:2009zg} etc. Moreover, a flat DA means
pion is an elementary field and can interact with quarks locally.
Thus, a quark loop can also give the logarithmic enhancement but
with rather low constituent quark
masses~\cite{Dorokhov:2009dg,Dorokhov:2010bz,Pham:2011zi}. In this
case, the QCD evolution is switched off, and the whole process is
totally non-factorizable~\cite{Radyushkin:2009zg}.  Therefore, a
rather flat DA with vanishing endpoints may still be a temporary
solution~\cite{Agaev:2010aq} taking into account that it is not
supported by the $u,d$ quark behavior in the $\eta$ and
$\eta^\prime$ TFFs~\cite{Kroll:2010bf}. It is even claimed that a
peculiar mechanism beyond the standard model is needed to explain
the BaBar puzzle~\cite{Roberts:2010rn,Brodsky:2011,Bakulev:2011rp}.

In the chiral limit approximation, there is only one large physical
scale in the hard part of pQCD calculation. So the power law is
explicit and concrete. In turn, the abnormal growth of the rescaled
pion TFF implies that an additional mass scale would be present. A
natural mechanism for generating a mass scale should satisfy the
following constraints: i) It should keep the general OPE analysis.
In another word, the solution should originate from non-OPE or
non-power corrections if only light quarks are involved; ii) The new
mechanism should leave the chiral anomaly intact at $Q^2=0$ since
the room for the theoretical and experimental improvement is very
small~\cite{Ioffe:2008tk}; iii) The pQCD asymptotic prediction
should be valid, though the current experimental region in the
exclusive processes seems to be still far away from the asymptotic
region determined in inclusive processes~\cite{Isgur:1984jm}; iv)
The flavor symmetry breaking of the DA should be explained in the
same scheme, and it is better to keep SU(3) flavor symmetry taking
into account its big success in the effective theory; v) Pion should
be treated as a bound state of current quarks. It is necessary to
make the new mechanism compatible with the numerous analyses for
both the asymptotic-like DA and collinear factorization before the
BaBar data since they have been cross checked based on different
powerful tools.

To find a solution that satisfies the above constraints
simultaneously seems not easy. In the literature there are also
proposals to suggest that the BaBar puzzle may need explanations
beyond QCD. However, before jump to any unusual solution, we should
first have a full understanding of the evolution of non-perturbative
mechanism from low to high $Q^2$. In principle, any solution
satisfying those constraints based on QCD should be treated
seriously.

In this work, we propose a novel prescription to address above
questions as a compensation of pQCD. Since non-perturbative flavor
changing processes (FCP) may play a role as ``mass
transmutation"~\cite{Polyakov:2009je} above certain momentum
transfers, we assume that such non-perturbative mechanisms can be
expressed in terms of hadronic degrees of freedom. With an effective
Lagrangian approach (ELA), we shall show that intermediate meson
loops have a peculiar evolution in terms of $Q^2$, which could be an
important mechanism causing the TFF deviations from the asymptotic
behavior in the region of $9<Q^2<40$ GeV$^2$.

BaBar Collaboration recently also measured the $\eta_c$
TFF~\cite{Lees:2010de}. A continuous growth of the rescaled TFF (or
slow decrease of the TFF) is observed clearly with the space-like
momentum transfer square up to $40 \ \mathrm{GeV}^2$, which is
caused by the large charm quark mass~\cite{Feldmann:1997te}. It is
also shown that a small charm component in $\eta^\prime$, which is
needed to explain the abnormally large branch ratio of $B\rightarrow
K\eta^\prime$ and $B\rightarrow
X_s\eta^\prime$~\cite{Halperin:1997as,Cheng:1997if}, may cause large
deviations from the asymptotic prediction in the medium momentum
transfer region while affects little the low momentum transfer
region~\cite{Feldmann:1997vc}.  So it is natural to speculate that
the charm quark may play a role in the deviations from the
asymptotic predictions in the BaBar kinematic region. The first
possible QCD diagram is the photons fuse into  a charm quark loop
which then couples to two gluons. These two gluons then evolve to
the pion through triangle anomaly. This factorizable charm loop is
also a possible correction for the chiral
anomaly~\cite{Ioffe:2008tk}, and cannot be recognized by the OPE
based only on the massless quarks. Although the correction at
$Q^2=0$, i.e. the chiral anomaly, is tiny, its evolution with the
photon virtuality could make it relatively important in larger
photon virtuality. Unfortunately, for the case of pion transition it
is an isospin suppressed process proportional to the $u,d$ quark
mass difference. Therefore, its influence on $\eta$ is much stronger
than that on the pion. To avoid the isospin suppression, the $u,d$
quarks would interact with photons more directly, and the gluon
exchanges between charm quark and light quark must be
non-perturbative to evade the strong coupling suppressions. Such a
scenario can be equivalent to the introduction of the intermediate
meson loops based on the quark-hadron duality argument. It will be
our focus in this work to investigate such a non-perturbative effect
on the pion TFF in the BaBar kinematic region.

The rest of this paper is organized as follows: The effective
Lagrangian approach for the intermediate meson loops is introduced
in Sec. II. Numerical results and discussions are presented in Sec.
III, and a brief summary is given in Sec. IV.

\section{Effective Lagrangian approach for the intermediate meson loops}

The effective Lagrangian for the $D^{(*)}$ mesons
$(D^{(*)0},D^{(*)+},D_s^{(*)+})$ couplings to light pseudoscalar
mesons~\cite{Cheng:2004ru} has the following expression:
\begin{eqnarray}
\mathcal{L}&= & -ig_{D^*D \mathcal{P}}(D^i\partial^\mu
\mathcal{P}_{ij}D^{*j\dag}_\mu-D_\mu^{*i}\partial^\mu
\mathcal{P}_{ij}D^{j\dag}) \nonumber\\
&&+\frac{1}{2}g_{D^*D^*\mathcal{P}}\epsilon_{\mu\nu\alpha\beta}D_i^{*\mu}\partial^\nu\mathcal{P}^{ij}
\stackrel{\leftrightarrow}{\partial^\alpha}D_j^{*\beta\dag} \ ,
\end{eqnarray}
where $\stackrel{\leftrightarrow}{\partial^\alpha}\equiv
\stackrel{\rightarrow}{\partial^\alpha}-\stackrel{\leftarrow}{\partial^\alpha}$,
and $\mathcal{P}$ denotes the pseudoscalar octet mesons
\begin{equation}
\mathcal{P} =\left(\begin{array}{ccc}
\frac{\pi^0}{\sqrt{2}}+\frac{\eta_8}{\sqrt{6}}      &   \pi^+                  & K^+\\
\pi^-                                             &
-\frac{\pi^0}{\sqrt{2}}+\frac{\eta_8}{\sqrt{6}}      &    K^0
\\K^-
&\bar{K^0}&-\sqrt{\frac{2}{3}}\eta_8
\end{array}
\right) \ .
\end{equation}
The corresponding Lagrangians  for the photon and $D^{(*)}$
couplings are
\begin{eqnarray}
\mathcal{L}_{DD\gamma}&=& i e A_\mu
D^-\stackrel{\leftrightarrow}{\partial^\mu}D^+ +ieA_\mu
D_s^-\stackrel{\leftrightarrow}{\partial^\mu}D_s^+\ ,
\\
\mathcal{L}_{D^*D\gamma}&=& \left\{\frac{e}{4}
g_{D^{*+}D^+\gamma}\epsilon^{\mu\nu\alpha\beta}F_{\mu\nu}D^{*+}_{\alpha\beta}D^-\right.\nonumber\\
&&+\frac{e}{4}
g_{D_s^{*+}D_s^+\gamma}\epsilon^{\mu\nu\alpha\beta}F_{\mu\nu}D^{*+}_{s\alpha\beta}D_s^-\nonumber\\
&&\left.+\frac{e}{4}
g_{D^{*0}D^0\gamma}\epsilon^{\mu\nu\alpha\beta}F_{\mu\nu}D^{*0}_{\alpha\beta}\bar{D^0}\right\}+h.c.
\ , \\
\mathcal{L}_{D^*D^*\gamma}&=& i e
A_\mu\left\{g^{\alpha\beta}D_\alpha^{*-}\stackrel{\leftrightarrow}{\partial^\mu}D_\beta^{*+}
+g^{\mu\beta}D_\alpha^{*-}\partial^\alpha D_\beta^{*+}\right.\nonumber\\
&&\left.-g^{\mu\alpha}\partial^\beta
D_\alpha^{*-}D_\beta^{*+}\right\}+i e
A_\mu\left\{g^{\alpha\beta}D_{s\alpha}^{*-}\stackrel{\leftrightarrow}{\partial^\mu}D_{s\beta}^{*+}
\right.\nonumber\\
&&\left. +g^{\mu\beta}D_{s\alpha}^{*-}\partial^\alpha
D_{s\beta}^{*+}-g^{\mu\alpha}\partial^\beta
D_{s\alpha}^{*-}D_{s\beta}^{*+}\right\}\ ,
\end{eqnarray}
where $F_{\mu\nu}\equiv
\partial_\mu A_\nu-\partial_\nu A_\mu$,
$D_{\mu\nu}^{*0,+}\equiv
\partial_\mu D_\nu^{*0,+}-\partial_\nu D_\mu^{*0,+}$,
and the vertices $D^0 D^0\gamma$ and $D^{*0}D^{*0}\gamma$ do not
exist.

The coupling constants $g_{D^{*+}D^+\gamma}$ and
$g_{D^{*0}D^0\gamma}$ are extracted from ~\cite{Chen:2010re}:
\begin{eqnarray}\label{eq:coupling}
&&g_{D^{*+}D^+\gamma}=-0.5 \
\mathrm{GeV}^{-1},\,\,g_{D^{*0}D^0\gamma}=2.0 \
\mathrm{GeV}^{-1},\nonumber\\
&&g_{D_s^{*+}D_s^+\gamma}=-0.3 \ \mathrm{GeV}^{-1},
\end{eqnarray}
where the signs and relative size can also be deduced from
constituent quark model, see also~\cite{Becirevic:2009xp} for the
lattice calculation. Similarly, the coupling constant $g_{D^*D
\mathcal{P}}\equiv g_{D^{*}D\pi}=17.9$~\cite{AA:2001}, and
$g_{D^*D^* \mathcal{P}}$ can be obtained from the heavy quark
effective theory
\begin{equation}\label{eq:HQET}
g_{D^*D^*\mathcal{P}}=\frac{g_{D^*D\mathcal{P}}}{\sqrt{M_{D^*}M_D}}
\ .
\end{equation}

\begin{figure}
\includegraphics[width=0.5\textwidth]{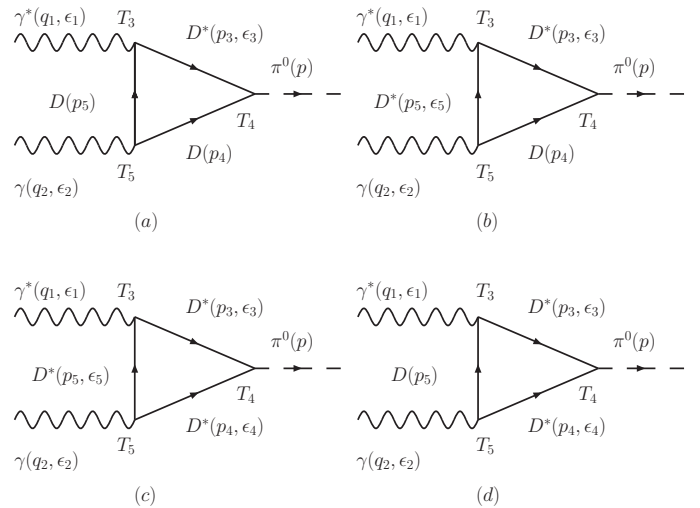}
\caption{$D^{(*)}$ meson loop contributions to pion TFF. The other
diagrams can be obtained by exchanging the photon momenta.}
\label{fig:1}
\end{figure}

Four types of loops would contribute to the pion TFF, i.e.
$D^*D(D)$, $D^*D(D^*)$, $D^*D^*(D^*)$, and $D^*D^*(D)$, where the
$D^{(*)}$ mesons in the parentheses are the exchanged mesons between
two photons. The kinematic conventions are illustrated by
Fig.~\ref{fig:1}. The transition amplitude can be expressed as
follows:
\begin{equation}
M_{fi}=\sum_{Polarization}\int\frac{\mathrm{d}^4p_3}{(2\pi)^4}
\frac{T_3T_4T_5}{a_3a_4a_5}F(p_3^2) \ ,
\end{equation}
where $T_{3,4,5}$ are the vertex functions given by the effective
Lagrangians, and $a_{i=3,4,5}$ are the denominators of propagators
of the intermediate mesons, respectively.

At hadronic level, the calculation of intermediate meson loops only
includes $S$-wave ground state mesons. In our case, the intermediate
charmed mesons include $D$ ($D_s$) and $D^*$ ($D_s^*$), and in the
strange sector $K$ and $K^*$ are considered. One essential question
which is very often raised is the contributions from higher excited
meson loops apart from those $S$-wave states. Intuitively, one would
expect that a complete set of intermediate meson loops should be
included based on the quark-hadron duality argument. In our
approach, the following reasons would allow us to only consider
contributions from the $S$-wave states. Firstly, for a given
kinematic condition, we would expect that the local EM dipole
couplings in Eq.~(\ref{eq:coupling}) would be much larger than the
excited states due to quantum mechanical selection rules. For
electric dipole couplings, e.g. $\gamma D^+D^-$, the ground state
would be less suppressed by the wavefunction overlaps than excited
states. Secondly, in many cases the quark-hadron duality is locally
broken such that a subset of hadron loops would play a dominant role
in observables. Moreover, we find in numerical simulations that
apart from the vertex couplings the loop integrals general decrease
when the masses of the intermediate mesons increase. In this sense
the contributions from the ground state $S$-wave intermediate meson
loops can be regarded as a reasonable approximation. Meanwhile,
although we still lack a systematic evaluation of how good such an
approximation would be in different circumstances, it interests us
to explore such a possible mechanism that accounts for the
non-factorizable contributions to the pion TFF.

We also note that in the process of $\gamma\gamma^*\to \pi^0$, there
exists a significant difference between $g_{D^{*+}D^+\gamma}$ and
$g_{D^{*0}D^0\gamma}$ as given in Eq.~(\ref{eq:coupling}). In some
other diagrams, the neutral meson loops even vanishes, e.g. due to
the $\gamma D^0\bar{D^0}$ vertex. It means that the isospin
violation effects are actually enhanced by the electromagnetic
interaction. Also, a nonvanishing charmed meson loop contribution
must be present in the low $Q^2$ region which comes from the
destructive sum between the charged and neutral $D^{(*)}$ meson
loops. At sufficiently high $Q^2$ the meson loop contributions
should become negligible as expected by pQCD. The interesting
question is at which range of $Q^2$ such an effect becomes
negligible. Meanwhile, an associated difficulty is to
model-independently quantify the loop contributions in a broad range
of $Q^2> 0$. In this sense our study is motivated to provide a
qualitative estimate of the intermediate meson loop effects as part
of the non-factorizable contributions in terms of $Q^2$.

Generally speaking,  the loops suffer from ultraviolet divergence
which means that we should replace the local coupling constants by
the non-local form factors to suppress the divergence. As usually
done, we introduce the form factor $F(p_3^2)$ to take into account
the off-shell effect of each vertex, which also reflects the
substructure of external hadrons. With the large space-like momentum
transfers, each vertex would be far off-shell. Therefore, we adopt
the following tri-monopole form in the calculation:
\begin{equation}\label{eq:formfactor}
F(p_3^2)=\prod_{i=3}^5\frac{(\Lambda_i^2-m_i^2)}
{(\Lambda_i^2-p_i^2)} \ .
\end{equation}
This turns out to be a reasonable consideration by our numerical
simulation. The cutoff $\Lambda_i$ can be parameterized as
$\Lambda_i=m_i+\alpha \Lambda_{QCD}$ with $\Lambda_{QCD}=220 \
\mathrm{MeV}$, and $m_i$ is the exchanged meson mass. Parameter
$\alpha$ is usually taken at $\mathcal{O}(1)$ and not necessarily
the same for different open flavors.

Apart from the $D$ loops, the Kaon and $B$ loops can also
contribute. The Kaon loops are similar to Fig.~\ref{fig:1}, where
the coupling constants can be determined by experimental
data~\cite{Nakamura:2010zzi}, i.e. $g_{K^{*+}K^+\gamma}=0.84 \
\mathrm{GeV}^{-1}$, $g_{K^{*0}K^0\gamma}=-1.27 \ \mathrm{GeV}^{-1}$,
$g_{K^{*+}K^+\pi^0}=-g_{K^{*0}K^0\pi^0}=-{9.17}/{\sqrt{2}}$,
$g_{\phi\eta\gamma}=0.69 \ \mathrm{GeV}^{-1}$ and
$g_{\phi\eta^\prime\gamma}=-0.73\ \mathrm{GeV}^{-1}$. The SU(3)$_F$
symmetry then gives
$g_{K^{*0}K^{*0}\pi^0}=-g_{K^{*+}K^{*+}\pi^0}=-{g_{\rho^0\omega\pi^0}}/{2}=-{11.58}/{2}\
\mathrm{GeV}^{-1}$.

\section{Results and discussions}

The introduction of an empirical form factor will bring in
model-dependent features. In order to give a meaningful
interpretation of the results, we outline the conditions for
constraining the parameter $\alpha$ as follows:

i) For an adopted value of $\alpha$, the meson loop corrections  to
the TFF at $Q^2=0$  should be negligibly small in order not to
conflict with the well established theoretical and experimental
results for the chiral anomaly~\cite{Ioffe:2008tk}. In practise, the
$D$ or Kaon meson loop contributions to the anomaly are restricted
to be less than $0.005\ \mathrm{GeV}^{-1}$ (absolute value) in
comparison with Eq.~(\ref{eq:inter}), $F^{\gamma\pi}(0)=0.27 \
\mathrm{GeV}^{-1}$. This serves as a strong constraint for the upper
limit of $\alpha$.

ii) The form factor of Eq.~(\ref{eq:formfactor}) introduces
additional singularities into the integrals
empirically~\cite{ff-loop}, which means that in order to reduce the
model-dependence, parameter $\alpha$ should have a sufficiently
large value, e.g. $\alpha > 1$. This condition can be satisfied in
$\gamma\gamma^*\to\pi^0$ since all the internal exchanged mesons are
highly off-shell.

iii) We neglect $B$ meson loops because the numerical calculation
shows that apart from couplings, their contributions are usually one
order magnitude smaller than $D$ and $K$ loops. In addition, we do
not include the non-strange light meson loops to avoid double
counting.

\begin{figure}
\includegraphics[width=0.5\textwidth]{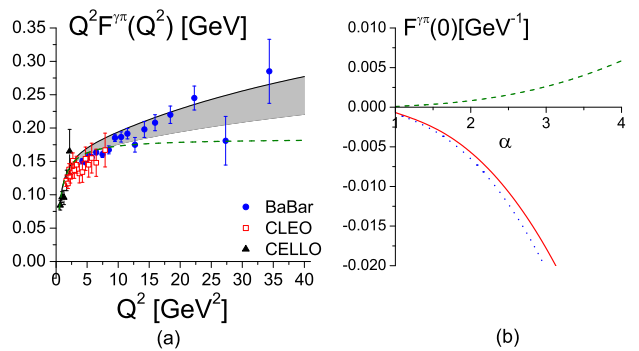}
\caption{(a) The rescaled pion TFF in comparison with the
experimental data. The dashed line is given by the pQCD
interpolation formula. The meson loops combining the pQCD results
give the shadowed band with $\alpha_D=2$ and $\alpha_D=3$ for the
lower and upper bound. (b) The $\alpha$ dependence of the meson loop
contributions to the chiral anomaly. The dashed, dotted, solid lines
are for $D$ loops, Kaon loops, and their sum with the same $\alpha$,
respectively.} \label{fig:3}
\end{figure}

In fact, we find that the above conditions can indeed provide a
stringent constraint on $\alpha$, and $1< \alpha_K < 2$, $1<
\alpha_D < 4$ can be determined in Fig.~\ref{fig:3}(b).

\begin{figure}
\includegraphics[width=0.5\textwidth]{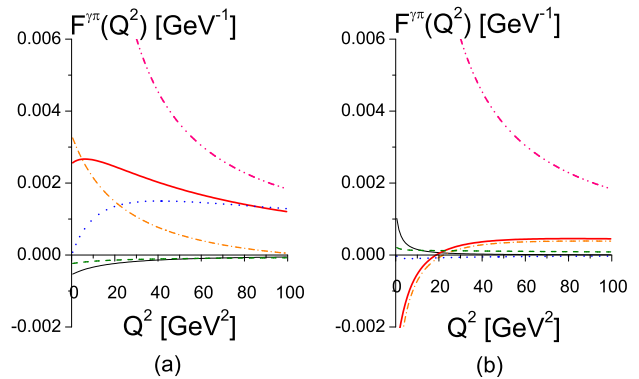}
\caption{(a) $D$ meson loop contributions to the pion TFF. The thin
solid, dashed, dot-dashed, and dotted lines denote $D^*D(D)$,
$D^*D(D^*)$, $D^*D^*(D^*)$, $D^*D^*(D)$ loops, and the thick solid
line for the total loops, dot-dot-dashed for Eq.~(\ref{eq:inter}).
(b) Kaon loop contributions with the same notations as (a), but
$D^{(*)}$ replaced by Kaons.} \label{fig:2}
\end{figure}

In Fig.~\ref{fig:2}, we plot pion TFF given by the exclusive $D$ and
Kaon loops with $\alpha_D=3$ and $\alpha_K=1.5$ in terms of $Q^2$.
It should be stressed that, the meson loop contributions to the
chiral anomaly are nearly zero because of the destructive
interferences from the $D$ and $K$ loops. It is indicated by the
thick solid line at zero momentum transfer point in
Fig.~\ref{fig:2}.  The dominance of $D$ loops over the Kaon loops
can be recognized. In particular, the $D^*D^*(D)$ loop exhibits an
enhancement in the BaBar kinematics due to the relatively large
coupling differences between the charged and neutral meson loops as
shown in Eq.~(\ref{eq:coupling}), i.e. the vertex coupling product
of the neutral $D$ meson loops is 16 times larger than that of the
charged $D$ mesons. The Kaon loops are much smaller because of the
destructive interferences among different loop amplitudes and
smaller $\alpha_K$.

A sensible feature of the $D$ loops is that, the total meson loop
contribution decreases faster than pQCD with the increasing $Q^2$,
Thus, it guarantees the asymptotic prediction of pQCD at high $Q^2$.
This is different from the logarithmic~\cite{Dorokhov:2010bz} or
double logarithmic~\cite{Dorokhov:2009dg,Pham:2011zi} growth
predicted by some analyses with flat DA or equivalent pion vertex
function, although we should caution that the ELA result at high
$Q^2$ is inevitably model-dependent.

In Fig.~\ref{fig:3}(a), the shadowed band is the exclusive
contributions from the intermediate meson loops plus the pQCD
prediction, which in this paper means the interpolation result of
Eq.~(\ref{eq:inter}). It should be noted that there is no double
counting between our meson loops and the pQCD interpolation result
involving only the $u,d$ quarks. The range of the shadowed area is
given by $\alpha_D=2$ (lower) and $3$ (upper bound) with
$\alpha_K=1.5$ fixed. This is a rather conservative estimate of the
meson loop contributions based on the above constraints on $\alpha$.
It shows that the inclusion of the meson loops significantly
improves the region of $Q^2 > 9$ GeV$^2$ while affects little in
$Q^2 < 5$ GeV$^2$. Therefore, the CELLO, CLEO and BaBar experiments
can be explained simultaneously without modifying conclusions
obtained before.  We also note that the discrepancies with the data
at low $Q^2$ are mainly caused by the pQCD interpolation formula
Eq.~(\ref{eq:inter}) which would break down due to the growing
importance of the QCD corrections with negative
sign~\cite{Mikhailov:2009kf}. We also emphasize that we do not try
to fit the data and judge the possible configuration of the pion DA.
The meson loop mechanism only tells us that in the high momentum
transfer region, the asymptotic-like pion DA may be still compatible
with the BaBar data.

\begin{figure}
\includegraphics[width=0.5\textwidth]{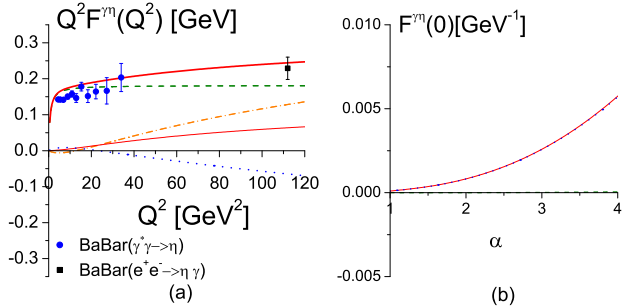}
\caption{(a) The rescaled $\eta$ TFF in comparison with the
experimental data.  The thick line denotes meson loops plus pQCD
interpolation result which is the dashed line, while the dot-dashed,
dotted and thin solid lines are for the $|\bar{n}n\rangle$,
$|\bar{s}s\rangle$, and physical $\eta$ from meson loops. (b) The
$\alpha$ dependence of the meson loop contributions to the chiral
anomaly. The dashed, dotted and solid lines are from the $D$ loops,
Kaon plus $\phi\phi(\eta^{(\prime)})$ loops, and their sum with the
same $\alpha$, respectively.} \label{fig:4}
\end{figure}

The meson loops can also contribute to the $\eta$ and $\eta^\prime$
TFFs. It is thus natural to expect that the same formulation in the
SU(3)$_F$ symmetry limit should keep consistent with the theoretical
and experimental results for $\eta$ and $\eta^\prime$, or at least
for $\eta$. Interestingly, the data~\cite{:2011hk} show that the
rise of the rescaled TFFs is much weaker than the rise of $\pi^0$
case. It implies that the $u,d$ components in $\eta$ and
$\eta^\prime$ would be significantly different from that in $\pi^0$
if the analysis is only based on pQCD~\cite{Kroll:2010bf}.
Therefore, it is interesting to examine the meson loop contributions
here as a direct test of the proposed mechanism. It should be
pointed out that a strict anomaly sum rule shows that the same
non-perturbative non-OPE correction to the continuum as in the pion
TFF should also be present in the $\eta$ and $\eta^\prime$
TFF~\cite{Klopot:2010ke,Klopot:2011ai}, although the $\eta$ TFF
favors the result from the pQCD interpolation formula. Here we do
not discuss the $\eta^\prime$ TFF because of its possible mixing
with the gluonic and/or charm component~\cite{arXiv:1110.6235}.

In Fig.~\ref{fig:4}, the rescaled $\eta$ TFF is plotted. Considering
the $\eta$-$\eta'$ mixing, an extended interpolation formula for
$\eta$ and $\eta'$ TFF was given by Foldmann and
Kroll~\cite{Feldmann:1998yc},
\begin{equation}\label{interp-eta}
F_{BL}^{\gamma P}(Q^2)=\frac{6C_q f_P^q}{Q^2+4\pi^2f_q^2}+\frac{6C_s
f_P^s}{Q^2+4\pi^2f_s^2} \ ,
\end{equation}
where $C_q=5/9\sqrt{2}$ and $C_s=1/9$, and $f_q$ and $f_s$ are decay
constants of the $\bar{q}q\equiv (\bar{q}q+\bar{d}d)/\sqrt{2}$ and
$\bar{s}s$ components. The following couplings are extracted in
Ref.~\cite{Feldmann:1998yc}: $f_q=1.07f_\pi$, $f_s=1.34f_\pi$ with
$f_\pi\equiv 0.131$ GeV, and the coupling transformation satisfies:
\begin{equation}
\left(
\begin{array}{cc}
f_\eta^q & f_\eta^s \\
f_{\eta'}^q & f_{\eta'}^s
\end{array}
\right) = U(\alpha_P)\left(
\begin{array}{cc}
f_q & 0 \\
0 & f_s
\end{array}
 \right) \ ,
\end{equation}
where the nonstrange $\bar{q}q$ and $\bar{s}s$ mixing angle favors a
range of $\alpha_P\equiv \theta_P+54.7^\circ\simeq 39^\circ \sim
43^\circ$~\cite{Feldmann:1998vh,hep-ph/0701020,Thomas:2007uy,:2011hk}.
With the above coupling constants and flavor mixing angle
$\alpha_P=39.3^\circ$ given by Ref.~\cite{Feldmann:1998yc}, one can
see clearly that the result of Eq.~(\ref{interp-eta}) does not
distinguish significantly from that of Eq.~(\ref{eq:inter}).
Remember that our intermediate meson loop contributions do not
interfere with the pQCD interpolation result. A further delicate
treatment of the interpolation formula can be regarded as an
improvement from the perturbative side, while our intermediate meson
loops are from the non-perturbative side.

In the meson loops the corresponding couplings (including $D_s$
loop) are $g_{D^*Dq\bar{q}}(q=u,d,s)=g_{D^*D\mathcal{P}}$ and
$g_{D^*D^*q\bar{q}}$ extracted from Eq.~(\ref{eq:HQET}). With
SU(3)$_F$ symmetry, there are $g_{K^*Kq\bar{q}}=-9.17$,
$g_{K^*K^*q\bar{q}}=11.58/\sqrt{2}\ \mathrm{GeV}^{-1}$, and
$g_{\phi\phi s\bar{s}}=11.58\times\sqrt{2}\ \mathrm{GeV}^{-1}$ for
the additional $\phi\phi(\eta^{(\prime)})$ loops. Including the
$\eta$-$\eta'$ mixing angle, the $\eta$ couplings to different
$D^{(*)}$ mesons can be obtained:
$g_{D^{*+}D^+\eta}=g_{D^*Dq\bar{q}}\cos\alpha_P /\sqrt{2}$,
$g_{D^{*0}D^0\eta}=g_{D^*Dq\bar{q}}\cos\alpha_P /\sqrt{2}$, and
$g_{D_s^{*}D_s\eta}=-g_{D^*Dq\bar{q}}\sin\alpha_P$.

The same form factor parameters, $\alpha_D=2.5$ and
$\alpha_K=\alpha_\phi=1.5$, as in the range of $\pi^0$ case are
adopted. Note that in Fig.~\ref{fig:4}(b), the constraints of
$\alpha$ are also satisfied by our setting. The total meson loop
correction is much weaker than in the pion case since the
$\bar{q}q\equiv (\bar{q}q+\bar{d}d)/\sqrt{2}$ and $\bar{s}s$
components significantly cancel out in $Q^2< 40$ GeV$^2$. Our result
suggests that the BaBar data can be well understood without
jeopardizing those old analyses of the $\eta$ TFF.

The datum at $Q^2=112 \ \mathrm{GeV}^2$ is actually measured at the
time-like point $Q^2=-112 \ \mathrm{GeV}^2$~\cite{Aubert:2006cy}.
Due to the analyticity of QCD, the time and space-like form factors
can be related to each other at $Q^2\to \infty$. We mark the
time-like datum point in Fig.~\ref{fig:4} assuming that the $|Q^2|$
value for the time-like virtual photon is high enough for the
analytical continuation of the TFF. In this sense, it may provide a
guidance for the magnitude of the corresponding space-like form
factor at $Q^2\simeq 112 \ \mathrm{GeV}^2$. It is interesting to see
that our result converges at high $Q^2$ as expected. Namely, the
meson loop contributions should vanish at high $Q^2$. Although the
pQCD interpolation formulas of $\eta$ and $\eta^\prime$ are not as
well-established as that of pion, e.g. radiative and higher-twist
corrections need to be systematically included, the role played by
the meson loops turns out to fit well the observed pattern.

\section{Summary}
In summary, we have proposed that the meson loops as a
non-perturbative component of the pion TFF may still play an
important role up to $Q^2\simeq 40$ GeV$^2$, hence cause deviations
from the collinear factorization results. This mechanism is
non-factorizable and different from the factorizable charm quark
loop at quark-gluon level. The  factorizable charm quark loop in the
pion TFF  must be suppressed by the $u$-$d$ quark mass difference as
compared with the $\eta$. The meson loops in terms of quark-hadron
duality may correspond to some continuum corrections in the spirits
of the strict anomaly sum rule~\cite{Klopot:2010ke}. It should be
stressed that this solution keeps consistent with the chiral anomaly
without violating the SU(3)$_F$ symmetry. Nevertheless, this
mechanism does not bring conflicts to against the pQCD analyses
before the BaBar result, and is an economic explanation for the
BaBar puzzle.

\section*{Acknowledgments}
Authors thank T. Huang, H.-N. Li, J.-W. Qiu for useful discussions.
This work is supported, in part, by the National Natural Science
Foundation of China (Grants No. 11035006), Chinese Academy of
Sciences (KJCX2-EW-N01), and Ministry of Science and Technology of
China (2009CB825200).

\end{document}